\documentclass[10pt,letterpaper]{article}
\usepackage[top=0.85in,left=2.75in,footskip=0.75in]{geometry}

\usepackage{amsmath,amssymb}
\usepackage{tikz}

\usepackage{changepage}

\usepackage{textcomp,marvosym}

\usepackage{cite}
\usepackage{pgfplots}
\usepackage{amssymb}
\usepackage{pifont}
\newcommand{\cmark}{\ding{51}}%
\newcommand{\xmark}{\ding{55}}%

\usepackage{nameref,hyperref}

\usepackage[nopatch=eqnum]{microtype}
\DisableLigatures[f]{encoding = *, family = * }

\usepackage{array}
\usepackage{tabularx,ragged2e,booktabs}

\newcolumntype{+}{!{\vrule width 2pt}}

\newlength\savedwidth

\newcommand\thickhline{\noalign{\global\savedwidth\arrayrulewidth\global\arrayrulewidth 2pt}%
\hline
\noalign{\global\arrayrulewidth\savedwidth}}

\raggedright
\usepackage[aboveskip=1pt,labelfont=bf,labelsep=period,justification=raggedright,singlelinecheck=off]{caption}

\makeatletter
\renewcommand{\@biblabel}[1]{\quad#1.}
\makeatother

\usepackage{lastpage,fancyhdr,graphicx}
\usepackage{epstopdf}
\fancyheadoffset[L]{2.25in}
\fancyfootoffset[L]{2.25in}
\begin{document}
\vspace*{0.2in}

\begin{flushleft}
{\Large
\textbf\newline{Context-Aware Stress Monitoring using Wearable and Mobile Technologies in Everyday Settings} 
}
\newline
\\
Seyed Amir Hossein Aqajari\textsuperscript{1*},
Sina Labbaf\textsuperscript{2},
Phuc Hoang Tran\textsuperscript{2},
Brenda Nguyen\textsuperscript{2},
Milad Asgari Mehrabadi\textsuperscript{1},
Marco Levorato\textsuperscript{2},
Nikil Dutt\textsuperscript{1,2,4},
Amir M. Rahmani\textsuperscript{1,2,3}
\\
\bigskip
\textbf{1} University of California, Irvine, Department of Electrical Engineering and Computer Science, Irvine, California, USA
\\
\textbf{2} University of California, Irvine, Department of Computer Science, Irvine, California, USA
\\
\textbf{3} University of California, Irvine, School of Nursing, Irvine, California, USA
\\
\textbf{4} University of California, Irvine, Department of Cognitive Sciences, Irvine, California, USA
\\
\bigskip

%
%



* saqajari@uci.edu

\end{flushleft}
\section*{Abstract}
\subsection*{Background}
Daily monitoring of stress is a critical component of maintaining optimal physical and mental health. 
Physiological signals and contextual information have recently emerged as promising indicators for detecting instances of heightened stress. Nonetheless, developing a real-time monitoring system that utilizes both physiological and contextual data to anticipate stress levels in everyday settings while also gathering stress labels from participants represents a significant challenge.

\subsection*{Objective}
We present a monitoring system that objectively tracks daily stress levels by utilizing both physiological and contextual data in a daily-life environment. Additionally, we have integrated a smart labeling approach to optimize the ecological momentary assessment (EMA) collection, which is required for building machine learning models for stress detection. We propose a three-tier Internet-of-Things-based system architecture to address the challenges. 

\subsection*{Methods}
A group of university students (n=11) consisting of both males (n=4) and females (n=7) with ages ranging from 18 to 37 years (Mean = 22.91, SD = 5.05) were recruited from the University of California, Irvine. During a period of two weeks, the students wore a smartwatch that continuously monitored their physiology and activity levels. A context-logging application was also installed on their smartphone. They were asked to respond to several EMAs daily through a smart EMA query system. We employed three different machine learning algorithms to evaluate the performance of our system. The mean decrease impurity approach was employed to identify the most significant features. The k-nearest neighbor imputation technique was used to fill out the missing contextual features.

\subsection*{Results}

 F1-score is the performance metric used in our study. We utilized a cross-validation technique to accurately estimate the performance of our stress models. We achieved the F1-score of 70\% with a Random Forest classifier using both PPG and contextual data, which is considered an acceptable score in models built for everyday settings. Whereas using PPG data alone, the highest F1-score achieved is approximately 56\%, emphasizing the significance of incorporating both PPG and contextual data in stress detection tasks.

\subsection*{Conclusion}

We proposed a system for monitoring daily-life stress using both physiological and smartphone data. The system includes a smart query module to capture high-quality labels. This is the first system to employ both physiology and context data for stress monitoring and to include a smart query system for capturing frequent self-reported data throughout the day.



\section*{Introduction}
Based on recent reports, a remarkable 70\% of individuals in the United States have encountered at least one symptom of stress within a given month \cite{bib1}.
Long-term stress can lead to a compromised immune system, cancer, cardiovascular disease, depression, diabetes, and substance addiction, among other serious effects \cite{bib1}.
In light of these consequences, the routine monitoring of stress levels has become increasingly essential.
Thus, developing dependable techniques for promptly detecting human stress is paramount.

The utilization of physiological signals as a modality for identifying stress has been extensively explored in the literature \cite{bib2, bib3}.
Among the physiological signals, the photoplethysmograph (PPG) signal is considered a valuable information source for stress detection \cite{bib4}.
This signal is influenced by the cardiac, vascular, and autonomic nervous systems, which are all known to be impacted by stress \cite{bib5}.
With the rapid development of wearable technologies, PPG signals can now be conveniently monitored in daily life settings using cost-effective wearable devices \cite{bib2}.
Moreover, the advancement of context-logging mobile applications has furnished a mechanism for continuously monitoring and tracking a user's contextual information, which encompasses location, activities, weather, and other pertinent factors, in real-time. Existing research has already illustrated the importance of this contextual information in comprehending and detecting stressful events experienced by individuals \cite{bib6}.

Real-time monitoring of physiological signals and contextual data presents a formidable challenge. The acquisition of physiological signals via smartwatches and wearable devices is particularly prone to motion artifact noise \cite{bib6.1}, necessitating extensive filtering and processing to enable their use in stress detection algorithms within daily life settings. Moreover, developing a daily life stress monitoring system mandates access to real-time stress level labels from participants, a task that poses several challenges \cite{bib6.2}. The timing of label querying is critical, requiring careful selection of moments when participants are not engaged in activities such as sleeping, studying, or working, to ensure optimal participation and reliable labels. Additionally, capturing the moments most conducive to experiencing stressful situations is crucial \cite{bib6.2}. Notwithstanding these obstacles, designing a robust and accurate system that captures both physiological and contextual information in real-time while querying labels from participants is an even greater challenge.

In this study, we present a context-aware daily life stress monitoring system that leverages physiological and contextual data and incorporates a smart label querying method. The system utilizes a publicly accessible life-logging mobile application to gather real-time contextual data from participants. The task of simultaneously collecting both contextual information and physiological signals while querying for stress labels in daily life presents significant challenges. To address these challenges, we propose a three-tier Internet of Things (IoT) based system architecture for our real-time monitoring system. 
In summary, the key contributions of this paper are as follows:
\begin{itemize}
    \item Propose a three-tier IoT-based system architecture to efficiently collect and record both physiological and contextual data alongside labels throughout the day.
    \item Implement a smart EMA triggering-based system to capture sufficient and high-quality labels multiple times daily.
    \item Investigate the impact of personalization on stress detection by examining how the performance of our algorithm improves with more subject-specific data available in the training phase.\end{itemize}


\section*{Related Works}
This section presents an overview of the related works as summarized in Table \ref{related-works}.
The majority of the existing research works in stress detection are conducted in laboratory settings or controlled environments \cite{bib7,bib8,bib9}.
In these studies, participants are typically required to wear wearable devices while engaging in a sequence of experimental tasks, such as viewing a series of images or videos or being exposed to stressful activities.
During the study, various kinds of bio-signals, such as PPG, Electrocardiogram (ECG), Electrodermal Activity (EDA), and Electroencephalogram (EEG) are recorded and employed for building models for stress detection.
Despite the remarkable performance obtained by these controlled experimental methods, such algorithms are not feasible for usage in real-world stress detection systems.
The data gathered in daily life is susceptible to contextual confounders and motion artifact noise resulting from movements and routine activities.
Moreover, the type of stress encountered in daily life scenarios can substantially differ from that induced in the controlled laboratory setting \cite{bib6.2}.

Recent advancements in stress detection methods have involved using physiological signals collected in real-world settings \cite{bib11,bib12,bib13,bib14,bib15}. This is achieved through the use of wearable devices, such as smartwatches or smart wristbands, which continuously collect physiological data from participants.  
Multiple questionnaires are sent randomly throughout the day to gather information on stress levels.
Finally, machine learning techniques and statistical algorithms are applied to the collected data to build a stress model.
A disadvantage of these studies is the absence of contextual information in their stress models, which can result in less reliable stress detection algorithms. The importance of contextual data in stress detection tasks has already been extensively demonstrated in the literature \cite{bib16}.

Can \textit{et al.} \cite{bib15} propose an objective stress detection system that uses smart bands and contextual information, such as weather information and activity type (e.g., lecture, presentation, or relaxation). However, one of the major limitations of this study is its semi-controlled setting. 
In their study, the data was obtained during an eight-day training event, where all the participants followed a predetermined schedule, including designated training days, free days, midterm presentations, and other similar activities.
Consequentially, the contextual data captured in this study was captured manually and is limited to the time and date of the predetermined schedule. 

Only a limited number of studies have investigated the integration of physiological signals and contextual information in a non-controlled, real-world setting \cite{bib17,bib18}.
However, the issue with these studies is the infrequency of their survey administration (i.e., once a day), as stress levels can vary greatly throughout the course of a day in response to various daily life events. These models cannot be properly coupled with mHealth-based just-in-time interventions due to their lack of assessing stress instantaneously. 
Increased frequency of survey administration would improve the likelihood of capturing fine-grained stress-inducing moments. 
Additionally, these studies lack a smart query system to capture the labels, which can result in more missing labels, such as instances where wearable devices are not being worn or carried. Our work aims to address these shortcomings using an objective and automatic physiological and contextual data collection approach focusing on fine-grained stress detection.

\begin{table}[!ht]
\begin{adjustwidth}{-2.25in}{0in} 
\centering
\caption{
{\bf Related Works}}
\begin{tabular}{|l|l|l|l|l|l|}
\hline
\multicolumn{1}{|l|}{\bf Study} & \multicolumn{1}{|l|}{\bf EMA Frequency} & \multicolumn{1}{|l|}{\bf Smart Query} & \multicolumn{1}{|l|}{\bf Physiology} & \multicolumn{1}{|l|}{\bf Context} & \multicolumn{1}{|l|}{\bf Daily Life}\\ \thickhline
Jeong Han et al. \cite{bib7} & N/A & \xmark & \cmark (PPG, ECG, SC) & \xmark & \xmark\\ \hline
Cho et al. \cite{bib9} & N/A & \xmark & \cmark (ECG) & \xmark & \xmark\\ \hline
Wang et al. \cite{bib16} & every 3 months & \xmark & \xmark & \cmark & \cmark\\ \hline
Yu et al. \cite{bib11} & 10/day & \xmark & \cmark (ECG, SC, ST, Motion) & \xmark & \cmark\\ \hline
Sah et al. \cite{bib12} & 4/day & \xmark & \cmark (PPG, SC, Motion, ST) & \xmark & \cmark\\ \hline
Tazarv et al. \cite{bib13} & distribution based & \cmark & \cmark (PPG) & \xmark & \cmark\\ \hline
Battalio et al. \cite{bib14} & 1/day & \xmark & \cmark (ECG, Motion, Resp) & \xmark & \cmark\\ \hline
Can et al. \cite{bib15} & 1/day & \xmark & \cmark (SC, PPG, ST, Motion) & \xmark & \xmark\\ \hline
Yu et al. \cite{bib17} & 1/day & \xmark & \cmark (SC, ST, Motion) & \cmark & \cmark\\ \hline
Mundnich et al. \cite{bib18} & 1/day & \xmark & \cmark (ECG, Motion) & \cmark & \cmark\\ \hline
\bf{This Work} & 7/day & \cmark & \cmark (PPG, Motion) & \cmark & \cmark\\ \hline
\end{tabular}
\label{related-works}
\end{adjustwidth}
\end{table}

\section*{Methods}
\subsection*{Study}
\label{study}

Starting in November of 2021, we recruited a sample of college students (n = 11) from the University of California, Irvine, via flyers and faculty outreach. The participants, comprising both male (n = 4) and female (n = 7) populations, ranged in age from 18 to 37 years (Mean = 22.91, SD = 5.05). The students were enrolled on a rolling basis at different intervals, depending on their enrollment date, and participated for a total of 2 weeks. During the enrollment process, participants review our study information document and are asked afterward if they agree to continue their participation. Consent is obtained verbally and is then documented in an Excel file by the research team member running the participant session.
As a component of the enrollment process, students were instructed to download 2 mobile applications (one foreground app to provide EMAs and one background app to perform passive mobile logging) and were equipped with a smartwatch.
Throughout the 2-week period, while wearing the smartwatch that continuously measured physiology and activity levels, students were prompted to complete multiple daily EMAs that was triggered by a smart EMA query system.

The experimental procedures involving human subjects described in this paper were approved by the Institutional Review Board (IRB) at the University of California, Irvine.

\subsection*{System Architecture}
\begin{figure}[t]
      \includegraphics[width=1\linewidth]{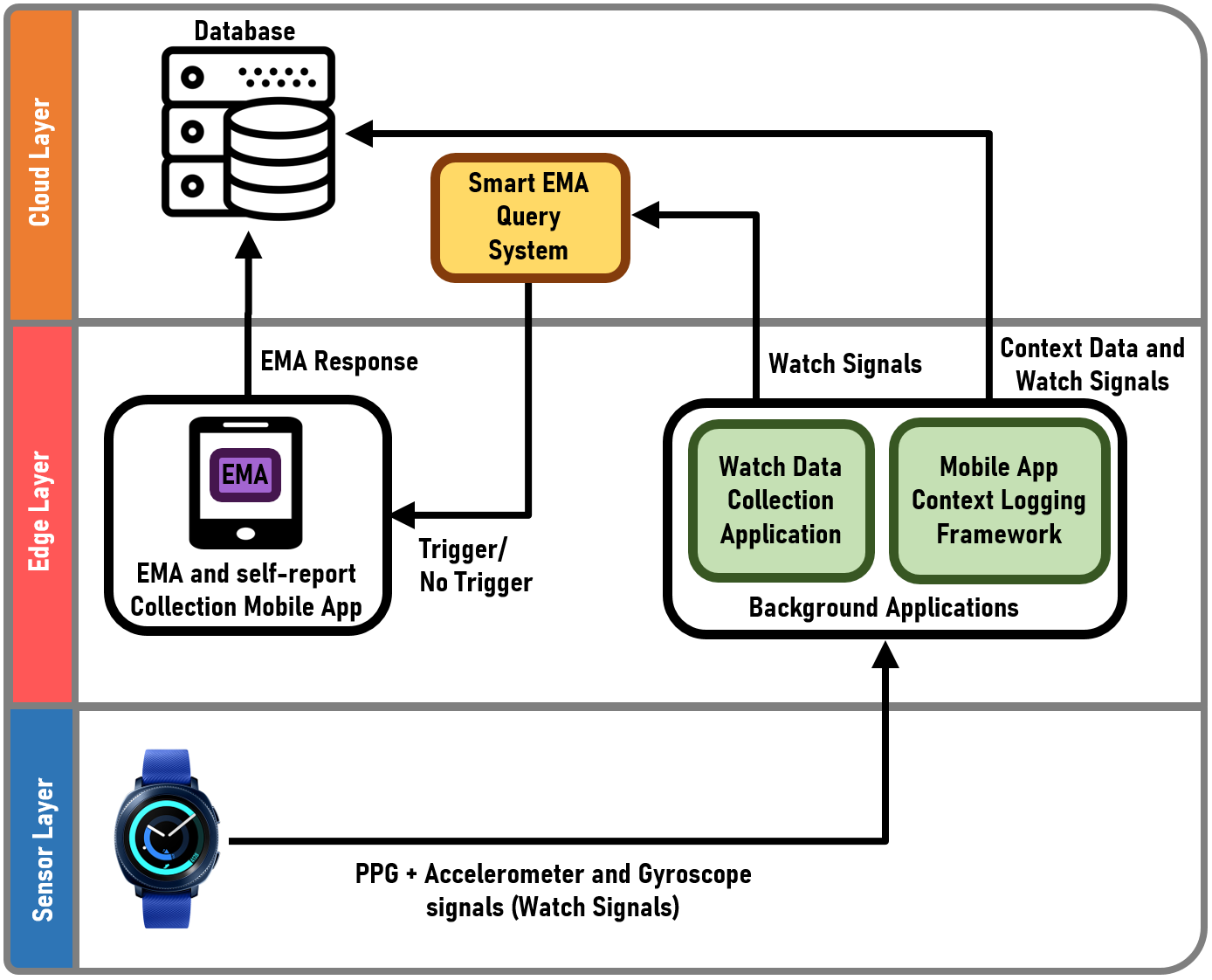}
      \caption{Proposed System Architecture}
      \label{fig:arch}
      \vspace{-1mm}
\end{figure}

The architecture of our proposed system is shown in Figure \ref{fig:arch}. 
The system comprises three primary layers that facilitate the collection of physiological and movement data, capturing contextual data, and querying labels.

\subsubsection*{Sensor Layer}
This study uses Samsung Galaxy Gear Sport Watches as the wearable device. 
This smartwatch is equipped with sensors capable of recording PPG (20Hz), accelerometer, and gyroscope (movement) signals.
We designed a custom smartwatch application for Samsung Galaxy Gear Sport Watches running on the Tizen operating system to gather these unprocessed PPG and movement signals. 
The data collected by the watch is transferred to the cloud layer when it is connected to a local Wi-Fi network, and in the absence of such a network, the data is transmitted via Bluetooth to a smartphone.
Two services and a user interface (UI) are included in the raw signal acquisition program. 
The initial service delivers sensor data to the cloud at intervals of two minutes which take place every fifteen minutes.
\subsubsection*{Edge Layer}
We use the AWARE framework \cite{bib19} to capture contextual data in everyday settings. 
AWARE is an open-source mobile instrumentation framework for logging, sharing, and reusing mobile context.
AWARE uses smartphone built-in sensors to capture daily life logging information such as phone battery level, weather, location, screen status, etc. 
In the event that the Wi-Fi network is inaccessible, an alternative smartphone application installed on the edge is employed to gather the raw PPG signals and accelerometer data from the sensor layer via Bluetooth and then transmit the data to the cloud for storage.
To elicit stress level labels from the study participants, a supplementary smartphone application has been developed which employs an EMA to solicit stress level ratings from the participants.

\subsubsection*{Cloud Layer}
A smart EMA query system is implemented (S-EMA) on the cloud to query labels throughout the day. 
The followings are the summary of the main rules used by the S-EMA module to trigger EMAs:
\begin{itemize}
    \item Sending EMAs only between 7 AM and midnight.
    \item Sending EMAs only when the user is wearing the watch using the accelerometer data.
    \item Sending EMAs only when the collected data is recent (the watch may record the data without the Internet and sync it later).
    \item It is intended to query labels seven times per day. The frequency of querying labels is adjusted dynamically to ensure that approximately seven labels are captured daily. The waiting period is calculated based on the initial wear time of the watch to achieve this target.
\end{itemize}
The stress levels in the EMAs are listed as ``not at all" (1), ``a little bit" (2), ``some" (3), ``a lot" (4), and ``extremely" (5).

Contextual data collected from the AWARE Framework (at the Edge layer), labels queried through EMAs (the Edge layer), and raw physiological (PPG) and movement (accelerometer, gyroscope, and gravity) signals captured by wearable platforms (the Sensory layer) are sent and stored in the cloud for cleaning, filtering, preprocessing, and being utilized in the predictive models.

\subsection*{Preprocessing}
The PPG signals stored in the cloud layer are collected from wearable devices and hence are prone to noise. To mitigate this issue, a number of preprocessing and filtering techniques are applied to the raw PPG signals in order to prepare them for further analysis.
To detect stress as a stimulus in  human subjects, a variety of features are extracted from these signals, such as heart rate, heart rate variability, and breathing rate, to name a few. 
Raw contextual data collected from the AWARE framework are also too broad and non-informative. 
A feature extraction module is designed with the purpose of transforming the raw contextual logging data into informative contextual life-logging features. These extracted features serve as inputs for our predictive machine-learning models.
Data Imputation and Feature Selection are two postprocessing techniques employed to improve our models' performance.

\subsubsection*{Data Cleaning and Windowing}
In order to clean the collected PPG data, first, we apply a Butterworth band-pass filter of order 3, with cut-off frequencies at 0.7Hz and 3.5 Hz. 
Then, a moving average across a 1-second window is  used  to  smooth  the  data  and  reduce  the artifacts  such  as  body gestures  and  movements,  which  are  common  in  everyday settings. 
These clean PPG signals, alongside contextual data collected from the AWARE framework, are resampled to 15-minute timing windows. 
Each of these 15-minute time frames, which consists of 2-minute continuous photoplethysmography (PPG) signals and context data collected through the AWARE framework, is then processed by the feature extraction module.

\subsubsection*{Feature Extraction}
In the feature extraction module, we use the HeartPy library \cite{bib20} to process the clean PPG signals to extract PPG peaks and PPG-relevant features including heart rate variability.
 The HeartPy is a Python Heart Rate Analysis Toolkit. 
 The toolkit is designed to handle (noisy) PPG data. 
 Using this library, the following 12 features are extracted from the PPG signals: BPM, IBI, SDNN, SDSD, RMSSD, PNN20, PNN50, HR\_mad, SD1, SD2, S, and BR. Table \ref{ppg-features} outlines the definitions of these features for reference.

 \begin{table}[!ht]
\begin{adjustwidth}{-2.25in}{0in} 
\centering
\caption{
{\bf PPG Features}}
\begin{tabular}{|l|l|}
\hline
\multicolumn{1}{|l|}{\bf Feature} & \multicolumn{1}{|l|}{\bf Definition}\\ \thickhline
BPM & Beats per minute, Heart Rate\\ \hline
IBI & Inter-Beat Interval, the average time interval between two successive heartbeats (NN intervals) \\ \hline
SDNN & Standard deviation of NN intervals\\ \hline
SDSD & Standard deviation of successive differences between adjacent NNs\\ \hline
RMSSD & Root mean square of successive differences between the adjacent NNs\\ \hline
PNN20 & The proportion of successive NNs greater than 20ms\\ \hline
PNN50 & The proportion of successive NNs greater than 50ms\\ \hline
HR\_mad & Median absolute deviation of NN intervals\\ \hline
SD1 and SD2 & Standard deviations the corresponding Poincare plot\\ \hline
S & Area of ellipse described by SD1 and SD2\\ \hline
BR & The number of breaths per minute (breathing rate)\\ \hline
\end{tabular}
\label{ppg-features}
\end{adjustwidth}
\end{table}

The raw contextual data captured from AWARE framework are presented in Table \ref{aware-features}.
These raw contextual data are not immediately usable in our predictive models and thus require further processing. To address this issue, we have implemented a feature extraction module to translate the raw data into numerical features that can be utilized by the models.

\begin{table}[!ht]
\begin{adjustwidth}{-2.25in}{0in} 
\centering
\caption{
{\bf Aware Features}}
\begin{tabular}{|l|l|l|l|}
\hline
\multicolumn{1}{|l|}{\bf Feature} & \multicolumn{1}{|l|}{\bf Definition} & \multicolumn{1}{|l|}{\bf Values} & \multicolumn{1}{|l|}{\bf Cut-offs}\\ \thickhline
battery\_adaptor & Indicator of power source & 0=No source,1=AC,2=Dock,3=USB & - \\ \hline
battery\_level & Battery percentage & (0:100]\% & [10,25,50] \\ \hline
speed & Movement speed of user &  Double value in m/s unit & [0,1,5]  \\ \hline
device\_off & Device not used duration & Double value in minutes unit & [2,10,20,60,180,540]\\ \hline
device\_on & Device is being used duration & Double value in minutes unit & [2,10,20] \\ \hline
air\_pressure & The ambient air pressure & Double value in mbar/hPa unit & [900,1000,1100]  \\ \hline
weather\_temperature & Measured temperature &  Double value in Celsius unit & [5,10,20,30] \\ \hline
weather & Weather forecast (API) & Weather forecast in Text, ex. 'Clear' & -\\ \hline
wind\_degrees & Degree of wind & Double value in degree  & [45,90,135] \\ \hline
wind\_speed & Speed of wind & Double value in m/s unit &  [0,2,5,10] \\ \hline
screen\_status & Status of phone screen & 0=off,1=on, 2=locked, 3=unlocked & -\\ \hline
location & Longitude, Latitude, Altitude & Three double values  & - \\ \hline
\end{tabular}
\label{aware-features}
\end{adjustwidth}
\end{table}

In Table \ref{aware-features}, the ``Values" column details the range, type, and units for each raw feature. The ``Cut-offs" column lists the threshold values utilized to convert the raw features into a more comprehensible and abstract numeric format suitable for usage in the predictive models. 
For example, the cut-off values for battery\_level are 10\%, 25\%, and 50\%. 
Therefore, for battery\_level (BL) values BL $\le10$,  $10<$ BL  $\le25$, $25<$ BL  $\le50$, and BL $>50$, we have respectively assigned the following numerical values: 0, 1, 2, 3. For battery\_adaptor and screen\_status features, there are no cut-off values, and their raw values are used in the model. 
For weather we used the following mapping function from text to the numerical values: \{'clear': 0, 'mist': 1, 'clouds': 2, 'rain': 3, 'snow': 4\}. 
In terms of location, since all participants are students at the University of California, Irvine (UCI), we abstracted and categorized their position into four distinct areas at the edge layer to preserve their privacy. 
These areas are defined as follows:
0: within the UCI recreation center (UCI ARC), 1: within the university premises (for work/study), 2: within UCI housing, and 3: outside of the aforementioned locations.
To assign a location to each participant, we  established circular boundaries encompassing each of these areas.
Once the Longitude, Latitude, and Altitude of the participant were determined, we checked whether their location lay within one of the circular boundaries. 
If it did, we would assign the corresponding numeric value; otherwise, the outside numeric value (3) would be assigned.

\subsubsection*{Data Labeling}
\begin{figure}[h]
\centering
\begin{tikzpicture}  

\begin{axis}  
[  
    ybar,  
    ylabel={\# Reported Labels}, 
    xlabel={\ Stress Levels (SL)},  
    bar width =20,
    symbolic x coords={SL1,SL2,SL3,SL4,SL5}, 
    xtick align=inside,
    grid=major,
    xtick=data,  
     nodes near coords, 
    nodes near coords align={vertical},  
    ]  
\addplot [draw=black,fill=gray]   coordinates { (SL1,288) (SL2,142) (SL3,120) (SL4,20) (SL5,21)};  
  
\end{axis}  

\end{tikzpicture} 
\caption{The distribution of reported stress levels}
\label{fig:lbl-dist}
\end{figure}
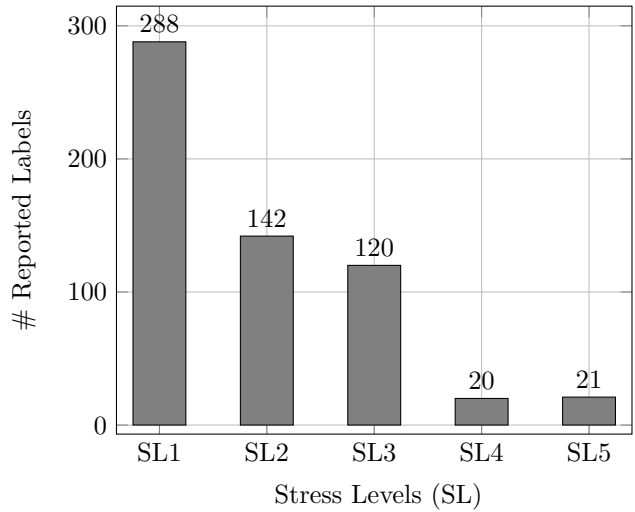

The EMA protocol is designed to trigger a maximum of seven times per day and prompt participants to indicate their stress level on a five-point Likert scale: (1) not at all, (2) a little bit, (3) some, (4) a lot, and (5) extremely.
The stress level reports, along with the corresponding timestamps, are recorded in the cloud for subsequent analysis. 
Each 15-minute timing window of collected physiological and contextual data is then labeled based on the closest subsequent EMA query.
The label distribution is shown in Figure \ref{fig:lbl-dist}.

\subsubsection*{Data Imputation}
The contextual features obtained from the AWARE framework are sourced from various sensors integrated within smartphones. 
As a result, certain 15-minute timing windows may exhibit missing data for some features, which can occur due to differences in the frequency of the sensors or technical limitations of a specific sensor. 
In order to optimize the construction of efficient machine learning models, it is recommended to utilize a data imputation technique rather than solely discarding incomplete data. This ensures the utilization of all available features. To this end, we use the k-nearest neighbor's imputation algorithm \cite{bib21} in order to compensate for missing contextual values in our data. 
The method predicts the values of any additional data points using "feature similarity." 
In other words, the value given to the new point depends on how much it resembles the points in the training set. 
Employing a technique that identifies the k-nearest neighbors to the observation with missing data, and subsequently imputes them based on the non-missing values in the vicinity, can be an advantageous approach for predicting absent values. This technique creates a rudimentary mean impute and then utilizes the resulting complete dataset to construct a KDTree. The KDTree is then utilized to compute the nearest neighbors (NN). Following the determination of k-NNs, it calculates their weighted average. This algorithm is deemed to be more accurate than the commonly utilized imputation methods such as mean, median, and mode, as it incorporates the similarity between the features into consideration.

\subsection*{Stress Detection}
We evaluate the efficacy of our proposed stress detection system through binary classification. 
In this classification, instances of ``no stress" (represented by a stress level of 1) are assigned a value of 0, while instances of ``a little bit," ``some," ``a lot," and ``extremely" (represented by stress levels greater than or equal to 2) are assigned a value of 1.

The main reason for classifying the samples into ``stress'' and ``no stress'' is to have a more balanced distribution of labels since some classes such as ``extremely'' and ``a lot'' are rare. As shown in Figure \ref{fig:lbl-dist}, with this categorization, there will be 288 samples with label 0 (``no stress'') and 303 samples with label 1 ``stress''.

\subsubsection*{Machine Learning Algorithms}
To build a stress detection algorithm, we used machine learning-based methods. 
Three classification techniques were employed, namely the K-Nearest-Neighbor \cite{bib22} with k values ranging from 1 to 20, Random Forest \cite{bib23} with a depth range of 1 to 10, and the XGBoost classifier \cite{bib24}. 
The K-Nearest-Neighbor technique predicts the outcome based on a majority vote using the k number of closest data points. 
The Random Forest classifier is an ensemble learning approach that employs averaging to increase predictive accuracy and reduce overfitting. 
It fits a number of decision tree classifiers to different subsamples of the dataset. 
The XGBoost is an efficacious open-source implementation of the gradient-boosted trees algorithm.
Gradient boosting is a supervised learning process that combines the predictions of a number of weaker, simpler models to predict a target variable properly. 

\subsubsection*{Feature Selection}

In this work, feature selection constitutes a pivotal stage, and its inclusion can significantly enhance the performance of our model. This is largely due to the inherent constraints associated with the AWARE features. Specifically, some features may exhibit consistent values over time, thereby rendering them less reliable and less critical for classification. Additionally, certain features may present significant quantities of missing data due to the challenges encountered in the collection of contextual data, such as sensor malfunction. Furthermore, the filtering and cleaning of the PPG signals, necessary to eliminate motion artifact noise, may result in the loss of critical information from the signals. As a consequence, the extraction of features from PPG signals may be rendered less dependable. It is, therefore, imperative to identify and eliminate these aforementioned features from the model to mitigate their adverse effects on the model's efficacy.

Given that a tree-based machine learning classification algorithm is employed in our prediction models, we have elected to adopt a tree-based feature selection algorithm.
Random forest classifiers offer the mean decrease impurity and mean decrease accuracy feature selection approaches. 
In this study, we chose features using a mean decrease impurity technique.
Gini importance is another name for mean decrease impurity. 
Random forest uses numerous different decision trees. 
Every node in the decision tree represents a condition on one of the qualities, and it is a model of decisions that resembles a tree. 
These nodes divide the data into two sets, with the goal of having the data with the same labels end up in the same set in the best case. 
The criterion used to determine the best condition for each node is impurity.
The total decrease in node impurity averaged over all ensemble trees is what is meant by "mean decrease impurity" for each feature.
This metric is used to order the features.

\subsubsection*{Performance Metrics}
In order to evaluate the performance of our stress monitoring system, we use F1-score as a quality metric. 
The F1-score is a measurement of a test's accuracy used in statistical analyses of binary categorization. 
It is derived from the test's precision, and recall, where precision is the proportion of ''true positive" results to ''all positive results," including those incorrectly identified as positive, and recall is the proportion of ''true positive" results to ''all samples that should have been identified as positive." 
In diagnostic binary classification, recall is also referred to as sensitivity, while precision is also referred to as positive predictive value. 
The F1-score is calculated as the weighted average of precision and recall:
$$
F1 = 2\times\frac{\text{precision}\times\text{recall}}{\text{precision}+\text{recall}}
$$

\section*{Results}
In our research, a cross-validation technique \cite{bib25} was utilized to evaluate the performance of our classification models. Cross-validation is a widely employed algorithm for accurately estimating the performance of a machine-learning model on unknown data. The process involves training a model using different subsets of the data and then testing the average accuracy of the remaining data. To assess the effectiveness of our research findings, we employed a 5-fold cross-validation method. To ensure that there is no overlap of user data in the train and test splits, the splits were created based on user IDs.

In order to ensure objectivity and prevent any potential biases, we adopted a fresh start approach for each iteration of the stress detection model. 
We disregarded any prior knowledge or information from previous stress models or the data from the current test users. 
The ultimate performance of the model was computed by calculating the mean of the individual stress models' performances generated.

The summary of the performance achieved for our stress assessment algorithm utilizing solely PPG data with a 5-fold cross-validation technique is presented in Table \ref{tbl:results-ppg}. The table shows that KNN with k=7 exhibited the best performance with an F1-score of 56. However, the result of 56 is not promising for binary classification.

\begin{table}[!ht]
\begin{adjustwidth}{-2.25in}{0in} 
\centering
\caption{
{\bf Validation accuracy of our stress assessment algorithm using only PPG data}}
\begin{tabular}{|l|l|l|}
\hline
\multicolumn{1}{|l|}{\bf Classifiers} & \multicolumn{1}{|l|}{\bf F1} & \multicolumn{1}{|l|}{\bf Selected Features} \\ \thickhline
Random Forest (depth=9) & 52 & SD1, IBI, S, SD2, BPM, SDSD\\ \hline
KNN (k=7) & 56 & SD1, IBI, S, SD2, BPM, SDSD\\ \hline
XGBoost & 51 & S, SD2, BPM, SDSD\\ \hline
\end{tabular}
\label{tbl:results-ppg}
\end{adjustwidth}
\end{table}

As a subsequent measure, we resolved to augment our model with contextual information to enhance its performance. The results of the stress assessment algorithm that incorporates both PPG and contextual data are summarized in Table \ref{tbl:results-multi}. This assessment was carried out using a 5-fold cross-validation technique. According to the table, the Random Forest model with a depth of 5, employing the top five features chosen by the GINI index algorithm, attained the best performance. This outcome indicates a 14\% increase in performance, underscoring the noteworthy role of contextual data in stress detection techniques.

\begin{table}[!ht]
\begin{adjustwidth}{-2.25in}{0in} 
\centering
\caption{
{\bf Validation accuracy of our stress assessment algorithm using both PPG and contextual data}}
\begin{tabular}{|l|l|l|l|}
\hline
\multicolumn{1}{|l|}{\bf Classifiers} & \multicolumn{1}{|l|}{\bf F1} & \multicolumn{1}{|l|}{\bf Selected Contextual Features} & \multicolumn{1}{|l|}{\bf Selected PPG Features}\\ \thickhline
Random Forest (d=3) & 70 & weather, wind\_speed, device\_off, location & BPM\\ \hline
KNN (k=9) & 62 & weather, wind\_speed, device\_off, location, speed & BPM, S, SD2, SDSD, SDNN, BR\\\hline
XGBoost & 64 & weather, wind\_speed, device\_off, location, speed & BPM, S, SD2, SDSD, SDNN, BR, IBI\\\hline
\end{tabular}
\label{tbl:results-multi}
\end{adjustwidth}
\end{table}

The most important contextual features, as determined by our analysis, are weather, wind\_speed, device\_off, and location. Regarding the PPG signal, beat per minute (BPM) was identified as the most relevant feature for stress detection. However, the performance of the KNN and XGBoost classifiers was found to be lower. For these two classifiers, the top 11 and 12 features were selected, respectively.

\subsection*{Explainability of the Model}
This section employs the stress detection model with the most optimal performance, which is the Random Forest model detailed in Table \ref{tbl:results-multi}. This model employs the five most important features, namely bpm, weather, wind\_speed, device\_off, and location, for the predictions.

To explain how our machine learning model predicts stress in terms of extracted features, we use the SHAP method \cite{bib26}. 
SHAP (SHapley Additive exPlanations) is a game theoretic approach to explain the output of any machine learning model. 
It utilizes the traditional Shapley values from game theory and their related extensions to correlate optimal credit allocation with local explanations. 

The fundamental concept of Shapley value-based interpretations of machine learning models is to allocate credit for a model's output among its input features. While computing SHAP values can be quite intricate, as they are generally NP-hard, this is not the case for linear models which are more straightforward. In such cases, we can extract the SHAP values directly from a partial dependence plot. 
Having a prediction $f(x_i)$, the SHAP value for a particular feature $x_i$ is the difference between the anticipated model output and the partial dependence plot at the feature's value $x_i$.

\begin{figure}[h]
\begin{tikzpicture}
 
\begin{axis} [xbar,
              grid=major,
              bar width =25,
              symbolic y coords={device\_off,weather,wind\_speed,BPM,location},
              ytick=data,  
     nodes near coords]
\addplot [draw=black,fill=gray]    coordinates {
    (0.2,device\_off) 
    (0.15,weather) 
    (0.14,wind\_speed) 
    (0.12,BPM)
    (0.1,location)
};
\end{axis}
 
\end{tikzpicture}
\caption{SHAP mean absolute values}
\label{fig:shap-values}
\end{figure}
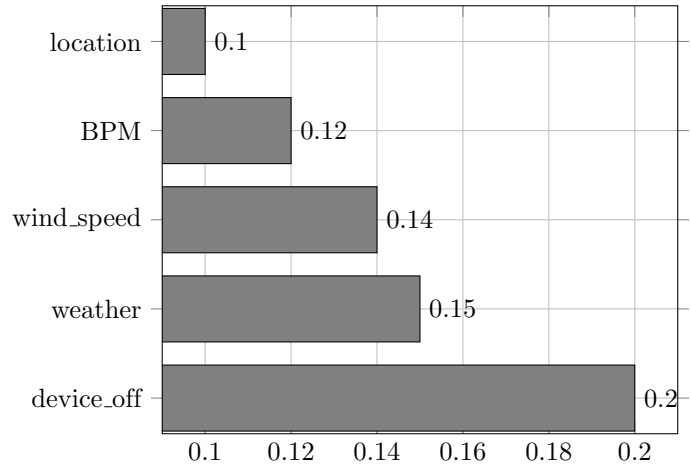

We apply SHAP method to our proposed stress detection model with the best performance, which is the Random Forest model presented in Table \ref{tbl:results-multi}. 
Figure \ref{fig:shap-values} shows the bar plot providing the absolute SHAP values calculated for each feature. 
This bar plot takes the mean absolute value of each feature over all the instances (rows) of the dataset (test data). 
According to this Figure, the BPM and location have the lowest impact on the model compared to the other features. 
The device\_off feature, which denotes the duration the phone is not in use, is a contextual feature that has the greatest impact on the model's outcome. Subsequently, the weather and wind\_speed features exhibit the highest influence on the model output, with a mean absolute SHAP value of approximately 0.15. 

\begin{figure}[t]
      \includegraphics[width=1\linewidth]{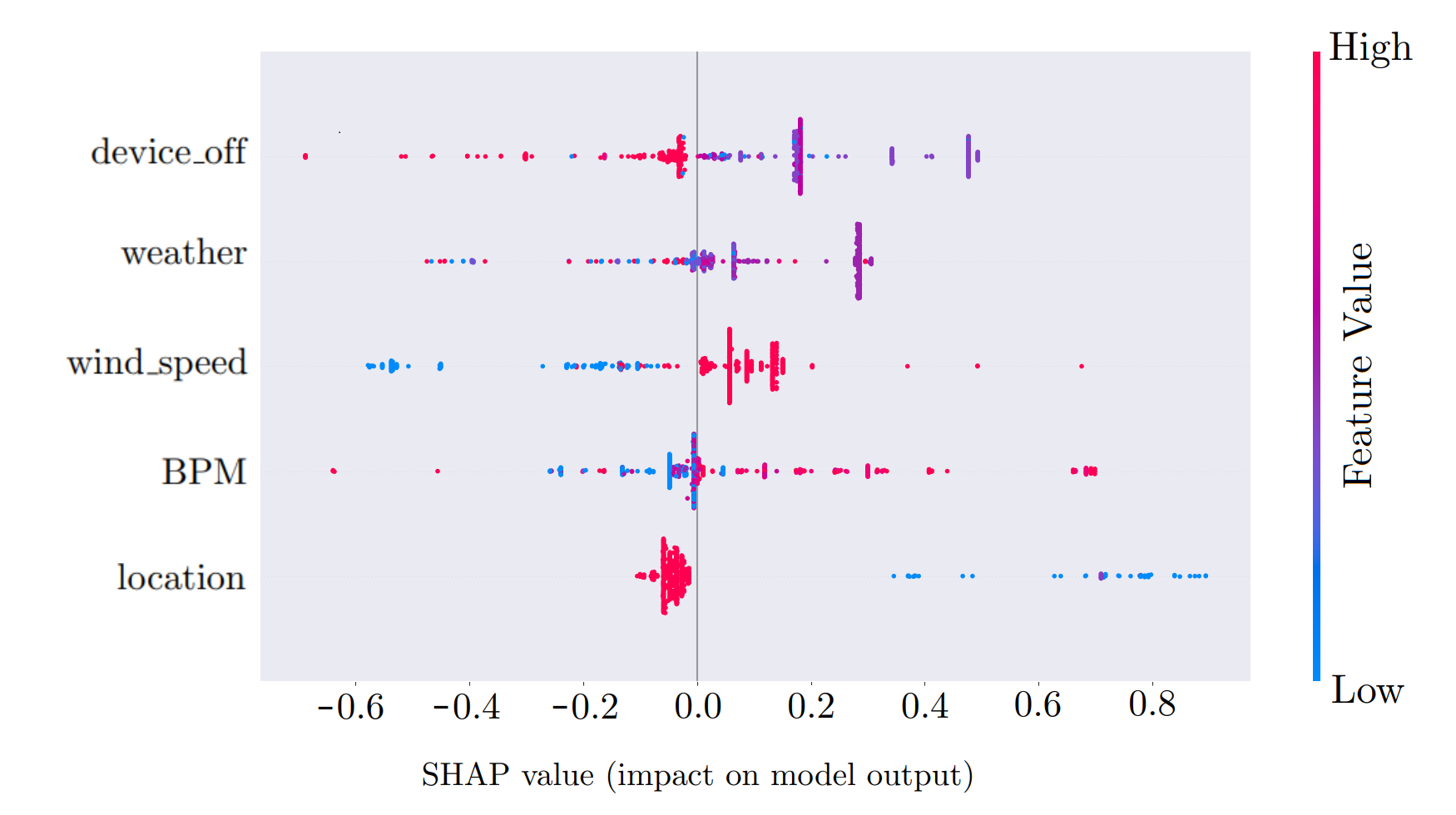}
      \caption{Feature impacts on the model output}
      \label{fig:impacts}
      \vspace{-1mm}
\end{figure}

To observe the impact of each feature on the model's output based on the feature values, we employ the beeswarm plot. Figure \ref{fig:impacts} presents a beeswarm plot that summarizes the complete distribution of SHAP values for each feature. Utilizing SHAP values, this plot showcases the effects of each feature on the model output. Features are sorted by the total sum of SHAP value magnitudes over all samples. The color of the plot demonstrates the feature value, with red indicating high and blue indicating low. This analysis shows that a high value for the device\_off feature (more time the device is not in use) results in a lower predicted stress value. As expected, a higher BPM increases the predicted stress value. For the weather feature, our findings suggest that when the weather condition is in the mid-range, such as mist, clouds, or rain, it increases the probability of stress, whereas when the weather condition is clear, it reduces the likelihood of stress. Furthermore, higher wind speed also increases the predicted stress value.

One notable finding that has been made here is that a lower value for the location feature indicates a higher predicted stress level while higher values indicate lower predicted stress levels. According to our designated ranges for the location feature, a lower value corresponds to the presence inside the university premises. On the other hand, higher values of the location feature correspond to UCI housing or an outdoor location, which results in decreased predicted stress levels. This observation suggests that the location of an individual can play a significant role in their stress levels, with certain locations associated with higher levels of stress.

\subsection*{Personalization}

In order to evaluate the impact of personalization on our stress detection algorithm, we conducted an experiment using data from three subjects with substantial amounts of data (S111, S912, and S731).
To achieve this, we trained our model on data from all other subjects in the first stage, and then tested it on half of the data from one of the selected subjects (e.g., S111).
In the second stage, we customized the model using the second half of the S111 data for training (in addition to data from other subjects) and the first half of the S111 data for testing.
We utilized the Random Forest classifier to demonstrate the performance changes.
Table \ref{tbl:pers_results} shows the results, indicating that personalization improves the prediction performance by approximately 10\%.
All the extracted PPG and contextual features were used in our models for this experiment.

\begin{table}[!ht]
\begin{adjustwidth}{-2.25in}{0in} 
\centering
\caption{
{\bf F1-score before and after personalization}}
\begin{tabular}{|l|l|l|l|}
\hline
\multicolumn{1}{|l|}{\bf User ID} & \multicolumn{1}{|l|}{\bf {F1 (Before)}} & \multicolumn{1}{|l|}{\bf {F1 (After)}} & \multicolumn{1}{|l|}{\bf Selected Features}\\ \thickhline
S111 & 74.1 & 81.0 & Random Forest (depth=5)\\ \hline
S912 & 76.0 & 83.0 & Random Forest (depth=5)\\ \hline
S731 & 36.7 & 47.1 & Random Forest (depth=19)\\ \hline
\end{tabular}
\label{tbl:pers_results}
\end{adjustwidth}
\end{table}

\section*{Discussion}
Capturing real-time features and signals while collecting labels from participants in daily life is challenging.
We propose a real-time monitoring system using a three-level architecture.
It includes a sensor layer with a Samsung Gear Sport watch 2, an edge layer with mobile apps using the AWARE framework, and a cloud layer to store data securely in a database.

Our real-time multi-tier system architecture was able to achieve an F1-score of 70\% for the task of stress detection.
In addition, we have shown that personalization has a positive impact on our stress detection models, resulting in  approximately a 10\% improvement.
This observation suggests that having a personalized model for the participants could result in improved performance for stress prediction models.

Despite the successful implementation of a multi-tier system, one limitation of our work is the occasional absence of contextual data for certain timing windows.
The contextual data is captured by the AWARE framework through different phone sensors, services, and APIs. 
Therefore, the limitations and potential inaccuracies of these sensors, services, and APIs, may result in missing features for some of the captured context data.

The second limitation of our research is associated with our label query system. Specifically, our labeling system functions on a time-event paradigm, wherein it solicits EMAs from study participants at pre-determined intervals of T hours. As a result, there exists a possibility that our system may not trigger an EMA request during instances when an individual is undergoing stress.
Conversely, the system may initiate an EMA  during moments when participants are resting or occupied with work, leading to an unsatisfactory experience and subsequently, an increase in missing EMA submissions.
In future work, we intend to implement a smarter query system to overcome the mentioned challenges with the purpose of (1) accurately identifying the time frames during which the participant experiences stress and (2) not resting or engaged in work or activities.

To efficiently identify these timing windows, it is imperative to establish a real-time data processing system in the cloud, which can receive and process data from the edge layer in real-time. By utilizing the processed physiological signals and contextual features, it becomes possible to detect circumstances that may result in stressful timing windows.
The accelerometer signals obtained from the Samsung Gear Sport Watch can be utilized to construct a machine-learning approach for identifying various daily activities, such as sleeping, walking, and sitting.
Daily-life activities alongside context features such as location could help us to build a smart high-level context recognition system detecting the most efficient timing windows to send the EMAs.

\section*{Conclusion}
In this work, we proposed a context-aware daily-life stress monitoring system using physiological and smartphone data. A smart query module, which uses accelerometer signals collected from a watch, is implemented in order to capture sufficient and high-quality labels. To the best of our knowledge, this is the first work presenting a daily-life stress monitoring system employing both physiology and context data with a smart query system to capture a sufficient number of EMAs throughout the day. According to our results, we were able to achieve an F1-score of up to 70\% using a Random Forest classifier. 

\section*{Acknowledgments}


%
%
%

\end{document}